\documentclass[12pt,notoc]{JHEP3} 

\usepackage{epsfig,multicol,bbm,amsmath}

\newcommand\fverb{\setbox\fverbbox=\hbox\bgroup\verb}
\newcommand\fverbdo{\egroup\medskip\noindent%
            \fbox{\unhbox\fverbbox}\ }
\newcommand\fverbit{\egroup\item[\fbox{\unhbox\fverbbox}]}
\newbox\fverbbox

\newcommand{\nablaslash}{\not{\hbox{\kern-3pt $\nabla$}}}

\title{Light-cone analysis of ungauged  and topologically gauged BLG theories}

\author{Bengt~E.W.~Nilsson\thanks{tfebn@chalmers.se} \\
    Fundamental Physics\\
Chalmers University of Technology\\
SE-412 96 G\"oteborg, Sweden\\}

\abstract{We consider three-dimensional maximally superconformal
Bagger-Lambert-Gustavsson (BLG) theory and its topologically gauged
version (constructed recently in  \cite{Gran:2008qx}) in the
light-cone gauge. After eliminating the entire Chern-Simons gauge
field, the ungauged BLG light-cone theory looks more conventional
and, apart from the order of the interaction terms, resembles
$\mathcal N=4$ super-Yang-Mills theory in four dimensions. The
light-cone superspace version of the BLG theory is given at
quadratic order together with a suggested form for the quartic
terms. Some problems with constructing the sixth order interaction
terms are also discussed. In the topologically gauged case, we
analyze the field equations related to the three Chern-Simons type
terms of $\mathcal N=8$ conformal supergravity and discuss some of
the special features of this theory and its couplings to BLG. }


\begin{document}


\section{Introduction}

A three-dimensional maximally superconformal Chern-Simons theory was
recently constructed in
\cite{Bagger:2006sk,Gustavsson:2007vu,Bagger:2007jr}. This
Bagger-Lambert-Gustavsson  (BLG) theory was originally proposed to
describe multiple M2-branes an interpretation that soon, however,
met with a number of problems stemming from the algebraic structure
of the theory. The Lagrangian contains a four-indexed structure
constant defining a three-algebra which is known to have essentially
only one (finite-dimensional) realization, $A_4$, related to the Lie
algebra $so(4)$ \cite{Papadopoulos:2008sk,Gauntlett:2008uf}. This is
limiting the role of this theory to stacks of two M2-branes.

To describe stacks of more than two M2 branes there are basically
two different options discussed in the literature. By relaxing the
assumption of a positive definite metric on the algebra any Lie
algebra can be accommodated. However, a degenerate metric
\cite{Gran:2008vi} leads to field equations which cannot be
integrated to a Lagrangian unless the potentially dangerous modes
can be rendered harmless. Postulating that these are constant makes
a Lagrangian possible but also seems to alter the theory in a
non-trivial way. Similar conclusions probably apply also to the case
of Lorentzian metrics.

The second possibility to avoid the $A_4$ uniqueness result
mentioned above is to reduce the number of (manifest)
supersymmetries from the maximal $\mathcal N=8$ to $\mathcal N=6$ as
done in the work of Aharony, Bergman, Jafferis and Maldacena (ABJM)
\cite{Aharony:2008ug}. These theories are, e.g., known to exist for
gauge groups $U(N)\times U(N)$ where $N$ can be any positive
integer.

In four dimensions the corresponding maximally supersymmetric theory
(with 16 supercharges) is the  $\mathcal N=4$ super-Yang-Mills
theory. This is a conventional theory with standard kinetic terms
for all fields, and with interaction terms dictated by the conformal
invariance. Instead, in three dimensions the BLG superconformal
theory with 16 supercharges, now corresponding to $\mathcal N=8$,
has several unusual features. One is the appearance of
three-algebras (and perhaps generalized Jordan triple systems
\cite{Nilsson:2008kq}\footnote{ See also \cite{Bagger:2008se} for
the special properties of the $\mathcal N=6$ structure constants
used in this context.}) and their associated structure constants
that satisfy the so called fundamental identity. A second
distinguishing feature is the fact that the BLG gauge potential has
dynamics governed by a Chern-Simons term
\cite{Schwarz:2004yj,Bagger:2006sk,Gustavsson:2007vu,Bagger:2007jr,Bandres:2008vf}
in the action, and is therefore not an independent field on-shell. A
standard kinetic Yang-Mills term is of course in conflict with
conformal invariance in three dimensions and will not be part of
this theory. However, such non-conformal Yang-Mills theories will in
general flow to an infrared fix-point where interacting
superconformal theories become relevant.

It is interesting to note that if written in the light-cone gauge
these two theories, $\mathcal N=8$ BLG in three and $\mathcal N=4$
SYM in four dimensions
\cite{Brink:1982wv,Mandelstam:1982cb,Brink:1982pd}, tend to look a
bit more similar, but with the crucial difference that the power of
the interaction terms differ. (As we will see later there are some
differences also in the way the $\partial_-$ derivatives appear.)
One might still hope, however, that after collecting all fields in
these two theories into superfields, one will find that the
Lagrangians will share the essential features needed to conclude
that their quantum properties coincide, i.e., that the BLG theory is
ultra-violet finite to all orders in perturbation theory for the
same reasons as in the $\mathcal N=4$ SYM case
\cite{Brink:1982wv,Mandelstam:1982cb,Brink:1982pd}. One purpose of
this note is to start the development of such a light-cone
superspace formulation of superconformal theories in three
dimensions.

We will also use light-cone methods to investigate the physical
properties (degrees of freedom) of topologically gauged BLG (i.e.,
$\mathcal N=8$ superconformal supergravity coupled to BLG) recently
discussed in \cite{Gran:2008qx}. This supergravity theory consists
of one ordinary Chern-Simons term (containing one derivative)
related to the R-symmetry and two slightly unusual higher derivative
Chern-Simons terms for the spin connection (three derivatives) and
the Rarita-Schwinger field (two derivatives)
\cite{VanNieuwenhuizen:1985ff,Lindstrom:1989eg}. The physical mode
content of these latter two terms are therefore less obvious but can
be easily analyzed in the light-cone gauge as will be demonstrated
below.

This note is organized as follows. In section 2 we review the
ordinary ungauged BLG theory and discuss some of its light-cone
properties. While the complete light-cone theory is presented in
component form, the superspace formulation is constructed only at
quadratic order. We do also suggest a possible answer at quartic
order. However, lacking the answer in light-cone superspace for the
sixth order interactions means that a proof of finiteness using
these methods will be out of reach for the moment. The complications
that are the reason for this are discussed briefly. Section three
starts with a short review of the $\mathcal N=8$ superconformal
supergravity theory followed by a light-cone analysis of its degrees
of freedom and couplings to BLG matter. Section four contains some
additional comments and conclusions.

\section{The BLG theory}

This section contains a review of the BLG theory followed by a
derivation of its light-cone Lagrangian. We also discuss some issues
that arise when trying to rewrite this theory in $\mathcal N=8$
light-cone superspace.

\subsection{Review of the BLG theory}

The BLG theory contains three different fields, the two propagating
ones $X^{I}_{A}$ and $\Psi_{A}$, which are scalars and spinors,
respectively, on the M2-brane, and the auxiliary gauge field
$\tilde{A}^{A}_{\mu B}$. Here the indices A,B,.. are connected to
the three-algebra and a basis $T^{A}$, while the I,J,K,.. are
$so(8)$ vector indices; in addition the spinors transform under a
spinor representation of  $so(8)$ but the corresponding index is not
written explicitly. Indices $\mu, \nu..$, etc, are vector indices on
the flat M2-brane world volume.

For these fields one can write down supersymmetry transformation
rules and covariant field equations. This does not require
 a metric on the three algebra which
means that the structure constants should be written as
$f^{ABC}{}_D$ and consequently the \emph{fundamental identity} reads
\begin{equation}
f^{ABC}{}_G f^{EFG}{}_D = 3f^{EF[A}{}_G f^{BC]G}{}_D  \,, \label{FI}
\end{equation}
which has as an alternate form \cite{Gran:2008vi}
\begin{equation}
f^{[ABC}{}_G f^{E]FG}{}_D = 0 \,. \label{WFI}
\end{equation}

The construction of a Lagrangian forces us to introduce a metric on
the three algebra, and if one wants to describe more general Lie
algebras than $so(4)$,  this metric must be generate
\cite{Gran:2008vi} or indefinite
\cite{Bandres:2008kj,Gomis:2008be,Ezhuthachan:2008ch,Cecotti:2008qs}
which, however, may just constitute a reformulation of D2-brane
systems. We also need to introduce the gauge field $A_{\mu AB}$
which is related to the previously defined one as follows
\begin{equation}
\tilde{A}_{\mu B}^{A}=A_{\mu CD}f^{CDA}{}_{B}.
\end{equation}

We will not need to concern ourselves in this paper with questions
related to the exact form of the structure constants as long as they
are compatible with the existence of a Lagrangian. The BLG
Lagrangian is
 \cite{Bagger:2007jr}
\begin{eqnarray}
{\cal L} &=& -\frac{1}{2}(D_\mu X^{AI})(D^\mu X_A^I) + \frac{i}{2}
\bar \Psi^A \Gamma^\mu D_\mu \Psi_A + \frac{i}{4} \bar\Psi_B
\Gamma_{IJ} X_C^I X_D^J \Psi_A f^{ABCD}
\\
&&-V +\frac{1}{2}\varepsilon^{\mu\nu\lambda}\left( f^{ABCD}A_{\mu
AB}\partial_\nu A_{\lambda CD} + \frac{2}{3} f^{CDA}{}_G f^{EFGB}
A_{\mu AB}  A_{\nu CD} A_{\lambda EF}  \right) \,,\nonumber
\end{eqnarray}
where
\begin{equation}
V = \frac{1}{12} f^{ABCD}f^{EFG}{}_D X_A^I X_B^J X_C^K X_E^I X_F^J
X_G^K \,.
\end{equation}

In order to rewrite this action in the light-cone gauge we need to
be able to express all non-propagating fields in terms of
propagating ones. These expressions are obtained from the components
of the field equations that become algebraic once the possibility to
divide by $\partial_{-}$ becomes available. This is one of the
crucial features of the light-cone gauge. The field equations
derived from the action above are
\begin{eqnarray}
0 &=& \Gamma^\mu D_\mu \Psi_A + \frac{1}{2} \Gamma_{IJ} X_C^I X_D^J
\Psi_B f^{CDB}{}_A \,,
\nonumber\\
0 &=& D^2 X_A^I -\frac{i}{2}\bar\Psi_C \Gamma^I{}_J X_D^J \Psi_B
f^{CDB}{}_A + \frac{1}{2} f^{BCD}{}_A f^{EFG}{}_D X_B^J X_C^K X_E^I
X_F^J X_G^K \,,\nonumber
\\
0 &=& \tilde F_{\mu\nu}{}^B{}_A - \varepsilon_{\mu\nu\lambda} (X_C^J
D^\lambda X_D^J + \frac{i}{2} \bar \Psi_C \Gamma^\lambda \Psi_D)
f^{CDB}{}_A \,,\label{FE}
\end{eqnarray}
where the covariant derivative and field strength are defined by
\begin{eqnarray}
(D_\mu X)_A &=& \partial_\mu X_A -\tilde A_\mu{}^B{}_A X_B \,,
\nonumber\\
\tilde F_{\mu\nu}{}^B{}_A &=& 2 \left(\partial_{[\mu} \tilde
A_{\nu]}{}^B{}_A + \tilde A_{[\mu}{}^B{}_{|C|} \tilde
A_{\nu]}{}^C{}_A \right) \,. \label{tFdet}
\end{eqnarray}

We will also need the supersymmetry transformation rules in the
analysis below. They are
\begin{eqnarray}
\delta X^{I}_A&=& i\bar\epsilon \Gamma^I \Psi_A,\cr \delta\Psi_A&=&
D_{\mu}X^I_A \gamma^{\mu}\Gamma^I \epsilon -\tfrac{1}{6}X^I_B X^J_C
X^K_D \Gamma^{IJK}\epsilon f^{BCD}{}_A,\cr \delta \tilde
A_{\mu}{}^A{}_B&=& i\bar\epsilon \gamma_{\mu} \Gamma^I X_C^I \psi_D
f^{CDA}{}_B.
\end{eqnarray}

\subsection{The BLG action in the light-cone gauge}

The light-cone quantities used in this paper are all defined in
accordance with the light-cone coordinates
\begin{equation}
x^{+}:=1/\sqrt{2}(x^0+x^1),\;\;\;x^{-}:=1/\sqrt{2}(x^0-x^1),\;\;\;x:=x^2,
\end{equation}
which means that the Lorentzian scalar product is
\begin{equation}
-x^0y^0+x^1y^1+x^2y^2=-x^{+}y^{-}-x^{-}y^{+}+xy.
\end{equation}
We choose the light-cone gauge on the vector field
$A_{\mu}=(A_-,A_+,A)$ as
\begin{equation}
A_{-}=0.
\end{equation}

Implementing these definitions and the gauge condition in the
bosonic part of the BLG action, it becomes
\begin{eqnarray}
{\cal L} &=&  \tfrac{1}{2}X^{I}_{A}(-2\partial_{-}\partial_{+}
+\partial^2)
X^{I}_{A}-X^{I}_{A}\tilde{A}_{+}^{AB}\partial_{-}X^{I}_{B}
+X^{I}_{A}\tilde{A}^{AB}\partial
X^{I}_{B}+\tfrac{1}{2}X^{I}_{A}(\tilde{A}^2)^{AB} X^{I}_{B}\cr
&-&A_{+}^{AB}
\partial_{-}\tilde{A}_{AB}-V\,,
\end{eqnarray}
where one may note that the interactions in the Chern-Simons term
are being put to zero by the gauge choice. The next step is to solve
the field equations involving $F_{\mu\nu}$ and insert the solutions
back into the above expression. We find that the $F_{-+}$ component
gives
\begin{equation}
\tilde A_{+}^{AB}=-\frac{1}{\partial_{-}}(X_C^I {\partial}X_D^I
-X_C^I X_E^I \tilde A^{E}{}_{D}
+\frac{i}{2}\bar{\Psi}_C\gamma\Psi_D)f^{CDAB},
\end{equation}
while  $F_{-2}$ implies that (we drop the index $2$ in the
following)
\begin{equation}
\tilde A^{AB}=\frac{1}{\partial_{-}}(X_C^I
{\partial_{-}}X_D^I+\frac{i}{2}\bar{\Psi}_C\gamma_{-}\Psi_D
)f^{CDAB}.
\end{equation}
The equation involving the remaining component of the field
strength, $F_{+2}$, then becomes an identity which may also be seen
from the Bianchi identity by using it to solve for this last
component in terms of the other two. Here we emphasize that
$\partial_{-}$ always acts just on the field following directly
after it, while $\frac{1}{\partial_{-}}$ acts on the whole
expression in the parenthesis following it.

We see here that to express the two non-zero components of $A_{\mu}$
entirely in terms of matter fields we must also insert $A$ back into
$A_+$. This is not needed in the analysis of four-dimensional
$\mathcal N=4$ SYM since in that case $A$ represents independent
degrees of freedom. In the analysis of the M2 system performed here,
eliminating also $A$ is straightforward and will, e.g., not cause
any problems related to functional determinants arising when doing
this in a path integral. This will be shown explicitly below.
However, to rewrite the full Lagrangian in light-cone superspace is
complicated and we will present the result only at quadratic order
and suggest a possible answer at quartic order. This problem and
some related issues will be discussed in the next subsection.

To repeat this for the spinor field $\Psi$, we start from its field
equation given in the previous subsection and solve for half the
spinor. This is done as follows: split $\Psi$ into
$\Psi^{(+)}=P^+\Psi=\tfrac{1}{2}(1+\gamma^2)\Psi$ and
$\Psi^{(-)}=P^-\Psi=\tfrac{1}{2}(1-\gamma^2)\Psi$ and use the fact
that the field equation produces the two equations
\begin{equation}
-\gamma^2D_2\gamma^-\Psi_A^{(+)}+\gamma^-\gamma^+D_+\Psi_A^{(-)}+
\tfrac{1}{2}\Gamma_{IJ}\gamma^-\Psi_B^{(+)}X_C^IX_D^J f^{CDB}{}_A=0,
\end{equation}
and
\begin{equation}
-\gamma^2D_2\gamma^+\Psi_A^{(-)}+\gamma^+\gamma^-D_-\Psi_A^{(+)}+
\tfrac{1}{2}\Gamma_{IJ}\gamma^+\Psi_B^{(-)}X_C^IX_D^J f^{CDB}{}_A=0,
\end{equation}
to solve for $\Psi^{(+)}$. We find (from the latter equation
suppressing the index $2$)
\begin{equation}
\Psi^{(+)}_A=-\tfrac{1}{2}\partial_-^{-1}(\gamma\partial\gamma^+\Psi_A^{(-)}+\gamma\tilde
A_A{}^B\gamma^+\Psi_B^{(-)})+\tfrac{1}{4}\partial_-^{-1}\Gamma_{IJ}\gamma^+\Psi_B^{(-)}X_C^IX_D^J
f^{CDB}{}_A.
\end{equation}
Also in this expression we need to insert $\tilde A^{AB}$ given in
terms of the matter fields above. As seen explicitly below, this
will produce terms in the light-cone Lagrangian that are sixth order
in fermionic fields.

We will now show in detail how to use the path integral to integrate
out all the dependent degrees of freedom, i.e., the remaining two
components of the Chern-Simons vector field and half of the BLG
spinor. From the form of the bosonic part of the action in the
light-cone gauge presented above, we see that the path integral is
(suppressing the R-symmetry indices and leaving out the fermions for
the moment to simplify the argument)
\begin{eqnarray}
Z[X]&=&\int {\cal D}[A,A_+]exp\,\,i\int
d^3x(-A_{+}^{AB}\partial_-\tilde A^{AB}\cr&&-\tfrac{1}{2}\tilde
A^{AB}(X^{[A}X_{[C})\delta^{B]}_{D]}\tilde A^{CD}+\tilde
A^{AB}(X_A\partial X_B)-\tilde A_{+}^{AB}(X_A\partial_-X_B)),
\end{eqnarray}
where the parenthesis  $(XX)$  indicates a scalar product in the
R-symmetry vector indices. This can be written  as
\begin{eqnarray}
Z[X]&=&\int {\cal D}[{\cal A}]exp\,\,i\int d^3x(\tfrac{1}{2}{\cal
A}^TM{\cal A}+{\cal A}^T{\cal J})
\end{eqnarray}
where, if we use the definitions
\begin{eqnarray}
{\cal A}^T=(\hat A,\tilde A)=(\partial_-A_+,\tilde A)
\end{eqnarray}
and
\begin{eqnarray}
{({\cal J}^T})^{AB}=( {J}^{AB},\tilde {J}^{AB})=
(f^{ABCD}\tfrac{1}{\partial_-}(X_C\partial_-X_D),(X^{[A}\partial
X^{B]})),
\end{eqnarray}
we find that the matrix $M$ is
\begin{eqnarray}
M_{AB}{}^{CD}&=& \left(
\begin{array}{ccc}
  & 0  &\delta_{AB}^{CD}\\
  & \delta_{AB}^{CD}&  H_{AB}{}^{CD}
\end{array}
\right),
\end{eqnarray}
with  $H$ given by
\begin{eqnarray}
H_{AB}{}^{CD}=-\delta_{[A}{}^{[C}(X_{B]}{X}^{D]}).
\end{eqnarray}

Performing the path integral over the vector field components
$A,A_+$ gives a functional determinant that is just
$det(\partial_-)^{-\tfrac{1}{2}}$ and hence field independent.
Furthermore, the matrix $M$ will appear in the path integral through
its inverse
\begin{eqnarray}
(M^{-1})_{AB}{}^{CD}=\left(
\begin{array}{ccc}
  & -H_{AB}{}^{CD}  & \delta_{AB}^{CD}  \\
  & \delta_{AB}^{CD}  &  0 \\
\end{array}
\right).
\end{eqnarray}
Explicitly we find (neglecting the determinant)
\begin{eqnarray}
Z[X]=exp\,\,i\int d^3x\,(-\tfrac{1}{2}{{\cal J}^T}M^{-1}{\cal J}),
\end{eqnarray}
where the integrand in the exponent reads
\begin{eqnarray}
\tfrac{1}{2}{{\cal J}^T}M^{-1}{\cal
J}&=&\tfrac{1}{2}f^{ABCD}\tfrac{1}{\partial_-}(X_C\partial_-X_D)(X_{[A}{X}^{[A'})\delta_{B]}{}^{B']}f^{A'B'C'D'}
\tfrac{1}{\partial_-}(X_{C'}\partial_-X_{D'})
\cr&&+f^{ABCD}(X_{A}\partial
X_{B})\tfrac{1}{\partial_-}(X_C\partial_-X_D).
\end{eqnarray}
This result is of course the same as that obtained by solving the
field equations for the vector field and inserting it back into the
action.

Adding the fermionic Lagrangian\footnote{We are working with the
following conventions:
$\gamma^0=i\sigma^2,\,\,\gamma^1=\sigma^1,\,\,\gamma^2=\sigma^3$ and
the Dirac conjugate is $\bar\Psi=\Psi^{\dagger}(-\gamma^0)$.}
\begin{eqnarray}
L_{fermion}=\tfrac{i}{2}(\sqrt{2}\Psi^+\partial_-\Psi^++2\Psi^+(\partial+\tilde
A)\Psi^-+\sqrt{2}\Psi^-(\partial_++\tilde A_+)\Psi^-)
\end{eqnarray}
to the above discussion just changes the matrix $M$ to
\begin{eqnarray}
\hat M_{AB}{}^{CD}= \left(
\begin{array}{cccc}
  & 0  & \delta_{AB}^{CD}& 0\\
  & \delta_{AB}^{CD}&  H_{AB}{}^{CD}& -i\delta_{[A}^{C}\Psi_{B]}^{(-)} \\
  & 0  &  i\delta_{A}^{[C}\Psi^{D](-)} & i\sqrt2\delta_A^C\partial_-\\
\end{array}
\right),
\end{eqnarray}
with  the same $H$ as before. Note that even after including the
fermions this matrix has a nice inverse
\begin{eqnarray}
(\hat M^{-1})_{AB}{}^{CD}= \left(
\begin{array}{cccc}
  &  -\hat H_{AB}{}^{CD}  & \delta_{AB}^{CD}&
  \,\,\,\,-\tfrac{1}{\sqrt2}\delta_{[A}^{C}\Psi_{B]}^{(-)}\tfrac{1}{\partial_-} \\
  & \delta_{AB}^{CD}&0 & 0\\
  & \tfrac{1}{\sqrt2}\tfrac{1}{\partial_-}\delta_{A}^{[C}\Psi^{D](-)} & 0&
   -\tfrac{i}{\sqrt2}\delta_A^C\tfrac{1}{\partial_-} \\
\end{array}
\right),
\end{eqnarray}
where $\hat H$ is given below. It should be noted here, however,
that this matrix is a slightly more delicate operator than $\hat M$
since the inverse derivatives $\tfrac{1}{\partial_-}$ are defined to
act on everything to the right of it (in particular the current
${\cal J}$).

To obtain these matrices we have defined the fermion extended
quantities,
\begin{eqnarray}
 \hat{\cal A} &=& (\partial_-A_+^{AB},\,\,\tilde
 A^{AB},\,\,\Psi_A^{(+)}),
\end{eqnarray}
and the current
\begin{eqnarray}
  \hat{\cal J} &=&(\hat {J}^{AB},\tilde {J}^{AB},{ J}_A),
\end{eqnarray}
where the respective components are
\begin{eqnarray}
\hat {
J}^{AB}&=&f^{ABCD}\tfrac{1}{\partial_-}(X_C\partial_-X_D)-f^{ABCD}
\tfrac{i}{\sqrt2}\tfrac{1}{\partial_-}(\Psi^{(-)}_C\Psi^{(-)}_D),\\
\tilde{ J}^{AB}&=&X^{[A}\partial_-X^{B]},\\
{J}_A&=&i({\partial}\Psi^{(-)}_A)+
\tfrac{i}{2}\Gamma_{IJ}\Psi^{(-)}_BX^I_CX^J_Df_A{}^{BCD}.
\end{eqnarray}

Performing the $\Psi^{(+)}$ part of the path integral is equivalent
to replacing $H$ with its fermion corrected version
\begin{eqnarray}
\hat
H_{AB}{}^{CD}=-\delta_{[A}{}^{[C}(X_{B]}X^{D]})+\tfrac{i}{\sqrt2}
\delta_{[A}{}^{[C}\Psi_{B]}^{(-)}\tfrac{1}{\partial_-}\Psi^{D](-)}.
\end{eqnarray}
This can be seen by considering the superdeterminant
\begin{eqnarray}
sdet\left(
\begin{array}{ccc}
  & A & C\\
  &D& B\\
\end{array}
\right)=det^{-1}(B)det(A-CB^{-1}D)
\end{eqnarray}
where the expression in the second determinant corresponds to the
new $\hat H$.

Collecting the above results, the action takes the following form
when expressed in terms of only propagating degrees of freedom:
\begin{eqnarray}
 {\cal L}  &=&  \tfrac{1}{2}X^I_A\Box
 X^I_A+\tfrac{i}{\sqrt2}\Psi_A^{(-)}\partial_+\Psi_A^{(-)}-
 \tfrac{1}{2}\hat{\cal J}^T\hat M^{-1}\hat{\cal J}-V,
\end{eqnarray}
where
\begin{eqnarray} \label{Llightconegauge}
\tfrac{1}{2}\hat{\cal J}^{\dagger}\hat M^{-1}\hat{\cal
J}&=&-\tfrac{1}{2}\hat J_{AB}^{\dagger}\hat H^{AB,CD}\hat
J_{CD}+\tfrac{i}{2\sqrt2}J^{\dagger}_A\tfrac{1}{\partial_-}J^A\nonumber
\\&&+\hat J^{\dagger}_{AB}\tilde J^{AB}+\tfrac{1}{\sqrt2}\hat
J^{\dagger}_{AB}\Psi_B^{(-)}(\tfrac{1}{\partial_-}J_A).
\end{eqnarray}
These expressions have in some cases a more  complicated
$\tfrac{1}{\partial_-}$ structure than is known from, e.g., SYM in
four dimensions. To see this let us consider the first one of these
terms
\begin{eqnarray}
    &-&\tfrac{1}{2}\hat J_{AB}^{\dagger}\hat H^{AB,CD}\hat
J_{CD}   =
 \tfrac{1}{2}f^{ABCD}f_A{}^{FGH}\times \nonumber\\
 && (\tfrac{1}{\partial_-}(X_C\partial_-X_D-
\tfrac{i}{\sqrt2}\Psi_C^{(-)}\Psi_D^{(-)})(X_BX_F)\tfrac{1}{\partial_-}(X_G\partial_-X_H-
\tfrac{i}{\sqrt2}\Psi_G^{(-)}\Psi_H^{(-)})\nonumber\\
   &-&  \tfrac{i}{\sqrt2}\tfrac{1}{\partial_-}(X_C\partial_-X_D-
\tfrac{i}{\sqrt2}\Psi_C^{(-)}\Psi_D^{(-)})
\Psi_{B}^{(-)}\tfrac{1}{\partial_-}(\Psi_F^{(-)}\tfrac{1}{\partial_-}(X_G\partial_-X_H-
\tfrac{i}{\sqrt2}\Psi_G^{(-)}\Psi_H^{(-)}))),\nonumber\\
\end{eqnarray}
where the last term contains nested inverse derivatives instead of
simpler combinations like $\tfrac{1}{\partial_-^2}$. Despite the
intricate structure of these sixth order terms they must be the ones
that are required in order to promote the original $X^6$ term in the
BLG Lagrangian to light-cone superspace. One of the terms above is
purely fermionic with three inverse $\partial_-$ derivatives to give
it the correct dimension.

Note also that the fermionic kinetic term in the light-cone
Lagrangian should be
$-\tfrac{i}{2\sqrt2}\Psi_A^{(-)}\tfrac{\Box}{\partial_-}\Psi_A^{(-)}$
and that the missing piece comes from the second term on the right
hand side of (\ref{Llightconegauge}).

Finally we remark that to fit the fields into a superfield, as
further discussed in the next subsection, we need to decompose $X^I$
and $\Psi^{(-)}$ into representations of $SU(4)$ according to
$\tfrac{1}{2}X^IY^I=\tfrac{1}{2}(A\bar B+\bar A
B)+\tfrac{1}{4}C_{mn}\bar D^{mn}$ and
$\Psi^{(-)}\Psi^{\prime(-)}=\chi_m\bar\chi^{\prime
m}+\bar\chi^m\chi^{\prime}_{m}$. It is then a trivial exercise to
write out the whole action in terms of the $8+8$ independent
light-cone degrees of freedom $A,C_{mn},\chi_m$.

\subsection{Light-cone superspace}

The purpose of this subsection is to try to organize the propagating
degrees of freedom in such a way that they will all fit into one
single light-cone superfield. The prototype superfield is the one
previously used in the context of four-dimensional $\mathcal N=4$
super-Yang-Mills \cite{Brink:1982wv,Mandelstam:1982cb,Brink:1982pd}
to prove the all loop finiteness of that theory (which ultimately is
one of the goals also here\footnote{See, e.g.,
\cite{Kapustin:1994mt,DelCima:1997pb} for some results on the
renormalization properties of Chern-Simons theories in three
dimensions. The actual BLG theory is discussed at one loop in
\cite{Gustavsson:2008bf}.}). As already discussed briefly at the end
of the previous subsection, by breaking $SO(8)$ to $SU(4)\times
U(1)$ we can define an $SU(4)$ complex scalar $A$ and a complex
field $C_{mn}$, with $m,n,..$ each a $4$ of $SU(4)$. The latter
field, being antisymmetric and selfdual in the indices, is then
transforming in a $6$ of $SU(4)$. Explicitly we define (suppressing
the three-algebra indices)
\begin{equation}
\tfrac{1}{2}X^I  X^I
=A\bar{A}+\frac{1}{4}C_{mn}\bar{C}^{mn},\,\,\,\,\Psi\Psi'=\bar\chi^m\chi'_m+\chi_m\bar\chi'^m,
\end{equation}
where the spinor bilinear $\Psi\Psi$ is the scalar product in their
$SO(8)$ spinor indices. Note that $A$ need not be confused with
components of the gauge field since they are all eliminated at this
stage. One may now use these expressions, together with the
decomposition in the action and write it completely in terms of
light-cone variables. This should be a suitable starting point in
the search for a light-cone superspace formulation of the BLG
theory.

The superspace version of the action at quadratic order is
\begin{equation}
S_2=-2^{-7}\int d^3xd^4\theta d^4 \bar\theta (\Phi
\frac{\Box}{\partial_-^{2}}\bar\Phi).
\end{equation}
We now need to define the superfield $\Phi$ (i.e., relate it to the
component fields) and verify that this superspace action provides
the correct component expression\footnote{Of course, this becomes
non-trivial first when discussing the interaction terms.}. To do so
we have to identify the covariant derivatives $d_m$ that are related
to the linearly realized generators $q_m$ of the supersymmetry
algebra: $d_m$ is the $P_-$ projected part of the covariant
derivative $D_{\alpha}$ (here $\alpha$ is an $SO(2,1)$ spinor index)
and satisfies, after decomposing its $SO(8)$ spinor index into $4$
and $\bar 4$,
\begin{equation}
\{d_m,\bar d^n\}=2\sqrt{2}i\delta^n_m\partial_-.
\end{equation}

Using standard superspace techniques, we replace the $d^4\theta$ in
the superspace measure with the covariant derivative expression
$d^4=\frac{1}{4!}\epsilon^{mnpq}d_md_nd_pd_q$ (and similarly for its
complex conjugate) and evaluate it on a generic integrand. Imposing
that $\Phi$ is chiral, $\bar d^m\Phi=0$, we find that
\begin{eqnarray}
d^4\bar d^4(\Phi\bar\Phi)&=&d^4\Phi\bar
d^4\bar\Phi-\tfrac{1}{6}d_md_nd_p\Phi\bar d^m \bar d^n \bar d^p
(2\sqrt{2}i\partial_-)\bar\Phi-\tfrac{1}{2}d_md_n\Phi\bar d^m \bar
d^n (2\sqrt{2}i\partial_-)^2\bar\Phi\cr&+&d_m\Phi\bar
d^m(2\sqrt{2}i\partial_-)^3\bar\Phi+\Phi(2\sqrt{2}i\partial_-)^4\bar\Phi.
\end{eqnarray}
Note that so far only the supersymmetry algebra and the
(anti)chirality constraint have been used which means that the
superfields in this expression can be replaced by any kind of
composite (anti)chiral combinations of the basic superfield.

The next step is to use this superspace expression to derive the
component form of the quadratic Lagrangian. To do this we first use
the duality constraint
\begin{eqnarray}
d_md_n\Phi=\tfrac{1}{2}\epsilon_{mnpq}\bar d^p \bar d^q \bar\Phi
\end{eqnarray}
(implying for instance that
$d^4\Phi=-(2\sqrt{2}i\partial_-)^2\bar\Phi$) to arrive at
\begin{eqnarray}
d^4\bar
d^4(\Phi\bar\Phi)&=&2\Phi(2\sqrt{2}i\partial_-)^4\bar\Phi-\tfrac{1}{2}d_md_n\Phi\bar
d^m \bar d^n (2\sqrt{2}i\partial_-)^2\bar\Phi\cr&&+2d_m\Phi\bar
d^m(2\sqrt{2}i\partial_-)^3\bar\Phi,
\end{eqnarray}
where we have allowed also for integration by parts. This gives the
result
\begin{eqnarray}
d^4\bar
d^4(\Phi\tfrac{\Box}{\partial_-^2}\bar\Phi)&=&-2^7(\partial_-\Phi\Box\partial_-\bar\Phi-\tfrac{1}{2^5}d_md_n\Phi\Box\bar
d^m \bar d^n \bar\Phi\cr&&-\tfrac{i}{2\sqrt2}\partial_-d_m\Phi
\Box\bar d^m\bar\Phi).
\end{eqnarray}

Finally, inserting the values of the superfield and its derivatives
at $\theta=0$, i.e.,
\begin{eqnarray}
\Phi\vert_0=\tfrac{1}{\partial_-}A,\,\,d_m\Phi\vert_0=\sqrt2\tfrac{1}{\partial_-}\chi_m,\,\,d_md_n\Phi\vert_0=2\sqrt{2}iC_{mn},
\end{eqnarray}
into the superspace action above gives the wanted answer
\begin{eqnarray}
{\mathcal L}_2=A\Box\bar A+\tfrac{1}{4}C_{mn}\Box\bar
 C^{mn}-\tfrac{i}{\sqrt{2}}\bar\chi^m\tfrac{\Box}{\partial_-}\chi_m.
\end{eqnarray}

Next we turn to the superspace interaction terms. As explained in
\cite{Mandelstam:1982cb}, these terms should be derivable  by
utilizing the non-linearly realized Lorentz and supersymmetry
transformations (and the irreducibility of the light-cone
superfield). However, this will not be done here for reasons that
will become clear in the following. Instead we note that in the
interaction terms the superfield $\Phi$ appears only to fourth and
sixth power due to the structure of the component action. For
instance, the terms involving only the field $C_{mn}$ are
\begin{eqnarray}
{\mathcal L}\vert_C&=&\tfrac{1}{4}C^A_{mn}\Box\bar
C^{Amn}-\tfrac{1}{4}(C^A_{mn}\partial\bar
C^{Bmn})\tfrac{1}{\partial_-}(C^C_{pq}\partial_-\bar
C^{Dpq})f^{ABCD}\cr&-&
\tfrac{1}{16}\tfrac{1}{\partial_-}(C^C_{mn}\partial_-\bar
C^{Dmn})(C^A_{pq}\bar
C^{Bpq})\tfrac{1}{\partial_-}(C^G_{rs}\partial_-\bar
C^{Hrs})f^{CDAE}f^{GHBE}\cr&-&\tfrac{1}{96}(C^A_{mn}\bar
C^{Emn})(C^B_{pq}\bar C^{Fpq})(C^C_{rs}\bar
C^{Grs})f^{ABCD}f^{EFGD}.
\end{eqnarray}
Note that $C^A_{mn}\bar C^{Bmn}$ is symmetric in $A$ and $B$ due to
the duality constraint, while if a derivative is inserted between
the two fields it also has an antisymmetric piece.

Some interesting features of these pure $C$ terms emerge if we try
to express them in superspace. First, the quartic  $C$ term seems to
tell us that the corresponding superspace term is just
\begin{eqnarray}
S_4=-2^{-7}\int d^3xd^4\theta d^4 \bar\theta
\tfrac{1}{16}\left((\Phi^A\partial\Phi^B)
\frac{1}{\partial_-}(\bar\Phi^C\partial_-\bar\Phi^D)f^{ABCD}\,+\,c.c.\right)
\end{eqnarray}
due to the following facts: $\Phi^A$ has dimension $+1/2$, the whole
$\mathcal N=8$ superspace measure has dimension $-1$, and $C$ is not
accompanied by any derivatives in the superfield.

Secondly, the sixth order terms in $\Phi^A$ must also contain  some
$\partial_-$ derivatives and/or covariant derivatives $d_m$ (and its
complex conjugate). Terms in the superspace Lagrangian can be
constructed for any set (of total dimension -2) of these derivatives
and it may be that a combination of such terms is needed. However,
since the last $C$ term above has no $\partial_-$ at
all\footnote{Note that no such term exists constructed from only
$A's$ and $\bar A's$.}, it is hard to see how any option with
explicit $\partial_-$'s could be realized. Thus using explicit
covariant derivatives $d_m$ (and its complex conjugate) seems to be
the only possibility. Note that this issue does not arise for
$\mathcal N=4$ super-Yang-Mills in four dimensions where the
corresponding term is of order four in the complex superfield.

Another option is to consider the related  $\mathcal N=6$ ABJM
theories \cite{Aharony:2008ug}.  These might in fact have a more
natural light-cone superspace formulation based on a complex
dimensionless superfield (and its conjugate) whose first component
is the fermionic field divided by $\partial_-$. Note that for
$\mathcal N=6$ the total superspace measure is dimensionless just as
in $\mathcal N=4$ SYM in four dimensions. This could, e.g., mean
that one can do without explicit $d_m$'s in the construction of the
superspace Lagrangian also in this case.

\section{The topologically gauged BLG theory on the light-cone}

In this section we first review the $\mathcal N=8$ conformal
supergravity theory discussed recently in \cite{Gran:2008qx}. As
noticed already in \cite{Lindstrom:1989eg}, the Lagrangian consists
of three Chern-Simons-like terms, one for each one of the fields in
the on-shell theory, the dreibein, the Rarita-Schwinger, and the
R-symmetry gauge field. The number of derivatives in the
corresponding 'Chern-Simons' terms are three, two and one,
respectively. In the second part of this section these terms are
analyzed in the light-cone gauge which will make the absence of
physical degrees of freedom clear. It will also show how the
derivative structure can be compatible with superconformal
invariance in the matter sector.

\subsection{Review}

The off-shell field content of three-dimensional ${\mathcal N=8}$
conformal supergravity is \cite{Howe:1995zm}
\begin{equation}
e_{\mu}{}^{\alpha}\,\,[0],\,\,\chi_{\mu}^i{}\,\,[-1/2],\,\,B_{\mu}^{ij}\,\,[-1],
\,\,b_{ijkl}\,\,[-1],\,\,\rho_{ijk}\,\,[-3/2],\,\,c_{ijkl}\,\,[-2],
\end{equation}
where the conformal dimensions are given in the square
brackets\footnote{Here the index $i$ can be any of the three
eight-dimensional representations of $SO(8)$.}. It is possible to
construct an on-shell topological Lagrangian from a set of
Chern-Simons terms \cite{Gran:2008qx} (see also
\cite{Lindstrom:1989eg}) using only the three gauge fields of 'spin'
2, 3/2 and 1, i.e.,
$e_{\mu}{}^{\alpha}[0],\,\,\chi_{\mu}^i{}[-1/2],\,\,B_{\mu}^{ij}[-1]$.

 The Lagrangian is a generalization of the
$\mathcal N=1$ case derived in \cite{Deser:1982sw} (see also
\cite{VanNieuwenhuizen:1985ff}) and takes the form
\begin{eqnarray}
L&=&\frac{1}{2}\epsilon^{\mu\nu\rho}
Tr_{\alpha}(\tilde\omega_{\mu}\partial_{\nu}\tilde\omega_{\rho}+
\frac{2}{3}\tilde\omega_{\mu}\tilde\omega_{\nu}\tilde\omega_{\rho})
-\epsilon^{\mu\nu\rho}Tr_i
(B_{\mu}\partial_{\nu}B_{\rho}+\frac{2}{3}B_{\mu}B_{\nu}B_{\rho})\cr
&&-i e^{-1} \epsilon^{\alpha\mu\nu}\epsilon^{\beta\rho\sigma}(\tilde
D_{\mu}\bar{\chi}_{\nu}\gamma_{\beta}\gamma_{\alpha}\tilde
D_{\rho}\chi_{\sigma}),
\end{eqnarray}
where $\tilde\omega$ is the spin connection and the traces in the
first and second terms are over the vector representation of the
Lorentz group $SO(1,2)$ and the R-symmetry group $SO(8)$,
represented by indices $\alpha$ and $i$, respectively\footnote{To
conform with the original work reviewed in this section we keep the
notation $i$ although the index for an $SO(8)$ vector representation
was denoted $I$ in the previous section.}.

We will frequently use the notation \cite{Deser:1982sw}
\begin{equation}
f^{\mu}=\frac{1}{2}\epsilon^{\mu\nu\rho}\tilde D_{\nu}{\chi}_{\rho},
\end{equation}
which makes the Rarita-Schwinger term read
\begin{equation}
-4i\bar f^{\mu} \gamma_{\beta}\gamma_{\alpha}
f^{\nu}(e_{\mu}{}^{\alpha}e_{\nu}{}^{\beta}e^{-1}),
\end{equation}
where we have spelt out explicitly all dependence of the dreibein
that needs to be varied when checking supersymmetry.

The standard procedure to obtain local supersymmetry is to start by
adding Rarita-Schwinger terms to the dreibein-compatible $\omega$ in
order  to obtain a supercovariant version of it. That is, we define
\begin{equation}
\tilde\omega_{\mu \alpha\beta}=\omega_{\mu \alpha\beta}+K_{\mu
\alpha \beta},
\end{equation}
where
\begin{equation}
\omega_{\mu\alpha\beta}=\frac{1}{2}(\Omega_{\mu\alpha\beta}-\Omega_{\alpha\beta\mu}+\Omega_{\beta\mu\alpha}),
\end{equation}
with
\begin{equation}
\Omega_{\mu\nu\alpha}=\partial_{\mu}e_{\nu}{}^{\alpha}-\partial_{\nu}e_{\mu}^{\alpha},
\end{equation}
and contorsion given by
\begin{equation}
K_{\mu\alpha\beta}=-\frac{i}{2}(\chi_{\mu}\gamma_{\beta}\chi_{\alpha}-
\chi_{\mu}\gamma_{\alpha}\chi_{\beta}-\chi_{\alpha}\gamma_{\mu}\chi_{\beta}).
\end{equation}
This combination of spin connection and contorsion is
supercovariant, i.e., derivatives on the supersymmetry parameter
cancel out if $\tilde\omega_{\mu \alpha\beta}$ is varied under the
ordinary transformations of the dreibein and Rarita-Schwinger field:
\begin{equation}
\delta e_{\mu}{}^{\alpha}=i\bar\epsilon \gamma^{\alpha}\chi_{\mu},
\,\,\, \delta\chi_{\mu}= \tilde D_{\mu}\epsilon.
\end{equation}

The covariant derivative appearing in the Lagrangian and in the
variation of the Rarita-Schwinger field takes the following form
acting on a spinor
\begin{equation}
\tilde
D_{\mu}\epsilon=\partial_{\mu}\epsilon+\frac{1}{4}\tilde\omega_{\mu\alpha
\beta}\gamma^{\alpha \beta}\epsilon+ \frac{1}{4}B_{\mu
ij}\Gamma^{ij}\epsilon,
\end{equation}
that is, both the Lorentz $SO(1,2)$ and the R-symmetry $SO(8)$
groups are gauged.

As explicitly demonstrated in \cite{Gran:2008qx} the above
Lagrangian is $\mathcal N=8$ supersymmetric (up to a total
divergence) under the above transformations of the dreibein and the
Rarita-Schwinger field together with a transformation of the $SO(8)$
R-symmetry gauge field $B_{\mu ij}$ that will be determined in the
course of the calculation. This superconformal  $\mathcal N=8$
supergravity theory can then be coupled to the BLG theory as also
discussed in \cite{Gran:2008qx}.

It is convenient to introduce the dual $SO(8)$ R-symmetry and and
curvature fields (see \cite{Deser:1982sw})
 \begin{equation}
G^{*\mu}_{ij}=\frac{1}{2}\epsilon^{\mu\nu\rho}G_{\nu \rho ij},\,\,\,
\tilde
R^{*\mu}{}_{\alpha\beta}=\frac{1}{2}\epsilon^{\mu\nu\rho}\tilde
R_{\nu\rho\alpha\beta}
\end{equation}
and similarly for $\tilde\omega$, as well as the double and triple
duals
\begin{equation}
\tilde
R^{**\mu,\alpha}=\frac{1}{2}\epsilon^{\alpha\beta\gamma}\tilde
R^{*\mu}{}_{\beta\gamma}, \,\,\, \tilde
R^{***}_{\mu}=\frac{1}{2}\epsilon_{\mu\nu\alpha}\tilde
R^{**\nu,\alpha},
\end{equation}
where in the last expression only the contorsion part of the Riemann
tensor contributes. In fact, one can show that
\begin{equation}
\tilde{R}^{***}_{\mu}=i \bar\chi_{\nu} \gamma_{\mu}f^{\nu}.
\end{equation}
One also finds that that
\begin{equation}
\delta\tilde\omega^{*\alpha}_{\mu}=-2i(\bar\epsilon\gamma_{\mu}f^{\alpha}-
\frac{1}{2}e_{\mu}{}^{\alpha}\bar\epsilon\gamma_{\nu}f^{\nu}).
\end{equation}

Combining this result with the fact that the commutator of two
supercovariant derivatives, acting on a spinor, is
\begin{equation}
[\tilde D_{\mu},\tilde D_{\nu}]=\frac{1}{4}\tilde
R_{\mu\nu\alpha\beta}\gamma^{\alpha\beta}+
 \frac{1}{4}G_{\mu\nu ij}\Gamma^{ij},
\end{equation}
we find that  the symmetric part of $R^{**\mu,\alpha}$ cancels in
the supersymmetry variation of the dreibein and gravitino
Chern-Simons terms. Performing also the variation of the
Chern-Simons term for the $SO(8)$ gauge field we find that also
$G^{*\mu}_{ij}$ cancels provided we choose the variation of $B_{\mu
ij}$ to be
\begin{equation}
\delta B_{\mu}^{
ij}=-\frac{i}{2}\bar\epsilon\Gamma^{ij}\gamma_{\nu}\gamma_{\mu}f^{\nu}.
\end{equation}

Inserting these variations into $\delta L$ gives
\begin{eqnarray}
\delta L&=&\delta L_1 + \delta L_2 + \delta L_3 + \delta L_4,\cr
\delta L_1&=&
4\bar\epsilon(\gamma_{\alpha}\gamma_{\beta}f^{\alpha})\bar
f^{\mu}\gamma^{\beta}\chi_{\mu},\cr \delta L_2&=&8\bar
f^{\mu}(\gamma_{\alpha}\gamma_{\beta}f^{\alpha})(\bar\epsilon\gamma^{\beta}\chi_{\mu}
-\frac{1}{2}e_{\mu}{}^{\beta}\bar\epsilon\gamma^{\nu}\chi_{\nu}),
\cr \delta L_3&=&4(\bar
f^{\alpha}\gamma_{\beta}\gamma_{\alpha})\gamma_{\gamma}\chi_{\mu}\epsilon^{\beta\mu\nu}
(\bar\epsilon\gamma_{\nu}f^{\gamma}-\frac{1}{2}e_{\nu}{}^{\gamma}\bar\epsilon\gamma^{\rho}f_{\rho}),\cr
\delta L_4&=&-\frac{1}{2}(\bar
f^{\alpha}\gamma_{\beta}\gamma_{\alpha})\Gamma^{ij}\chi_{\mu}\epsilon^{\mu\beta\gamma}\bar\epsilon
\Gamma_{ij}(\gamma_{\delta}\gamma_{\gamma}f^{\delta})).
\end{eqnarray}

In order to show that the variation of the Lagrangian vanishes some
of the terms in the above expression must be rearranged by Fierz
transformations. By applying the Fierz transformations to $\delta
L_1$ and $\delta L_3$ above and expressing all terms so obtained in
the Fierz basis one can show, after some ${\mathcal N=1}$ Fierz
calculations, that they exactly cancel $\delta L_2$. This is the
result of Deser and Kay \cite{Deser:1982sw}.

It now becomes rather easy to establish that also for $\mathcal N=8$
the variation will vanish when $\delta L_4$ is included and use is
made of the full $\mathcal N=8$ Fierz identity for $SO(8)$ spinors
of the same chirality, i.e.,
\begin{eqnarray}
\bar A B \bar C D& =&-\frac{1}{16}(\bar A D \bar C B + \bar A
\gamma_{\alpha}D \bar C \gamma_{\alpha}B \cr &&-\frac{1}{2}\bar A
\Gamma^{ij}D \bar C \Gamma^{ij} B- \frac{1}{2}\bar A\gamma_{\alpha}
\Gamma^{ij} D \bar C\gamma_{\alpha} \Gamma^{ij} B\cr
&&+\frac{1}{48}\bar A  \Gamma^{ijkl}D \bar C  \Gamma^{ijkl}B+
\frac{1}{48}\bar A \gamma_{\alpha} \Gamma^{ijkl}D \bar
C\gamma_{\alpha}  \Gamma^{ijkl}B).
\end{eqnarray}

This theory is also locally scale invariant (denoted by an index
$\Delta$) and possesses $\mathcal N=8$ superconformal (shift)
symmetry (denoted by $S$) with the following transformation rules
(where $\phi$ is the local scale parameter and $\eta$ the local
shift parameter)
\begin{eqnarray}
\delta_{\Delta} e_{\mu}{}^{\alpha}&=&-\phi(x)e_{\mu}{}^{\alpha},\cr
\delta_{\Delta} \chi_{\mu}&=&-\tfrac{1}{2}\phi(x)\chi_{\mu},\cr
\delta_{\Delta} B_{\mu}^{ij}&=&0,
\end{eqnarray}
and
\begin{eqnarray}
\delta_S e_{\mu}{}^{\alpha}&=& 0,\cr \delta_S \chi_{\mu}&=&
\gamma_{\mu} \eta, \cr
 \delta_S B_{\mu}^{ij}&=& \tfrac{i}{2}\bar \eta \Gamma^{ij}
 \chi_{\mu}.
\end{eqnarray}

\subsection{Light-cone analysis}

In this subsection we will analyze the three Chern-Simons terms that
make up the $\mathcal N=8$ superconformal supergravity theory
reviewed above. In \cite{Gran:2008qx} the superconformal
supergravity theory was coupled to the ordinary (ungauged) BLG
theory under the assumption that the supergravity sector does not
add any new propagating degrees of freedom. This is, in fact, quite
a natural property to expect from a theory that consists of just
Chern-Simons terms. However, in the case of the $\mathcal N=8$
superconformal supergravity theory only one of the three
Chern-Simons terms is a conventional one-derivative term, namely the
one for the gauged $SO(8)$ R-symmetry. The other two terms, on the
other hand, are unusual due to their number of derivatives, three
and two for the gravitational and the Rarita-Schwinger one,
respectively. Despite these complications, in \cite{Horne:1988jf}
the Chern-Simons term constructed from the metric compatible spin
connection was rewritten as an $SO(2,3)$ Chern-Simons theory making
clear some of its topological properties.

Here we will instead use light-cone techniques and analyze all three
terms in the $\mathcal N=8$ superconformal supergravity theory in a
similar fashion. The light-cone treatment of the Chern-Simons term
for the gauged $SO(8)$ R-symmetry is of course exactly the same as
for the BLG vector field $\tilde A_{\mu b}^a$ discussed in section
two.

Before going into the details of the light-cone analysis of the
other two fields in the supergravity theory we note that also they
have two field components each off-shell after making use of the
superconformal gauge invariance. Although the two Chern-Simons terms
are much more complicated one might hope that also here these two
field components will be completely determined. As we will see
below, this is indeed the case but the situation is slightly more
interesting than that.

Next we turn to the Rarita-Schwinger (or gravitino) term
\begin{eqnarray}
L=-i e^{-1} \epsilon^{\alpha\mu\nu}\epsilon^{\beta\rho\sigma}(\tilde
D_{\mu}\bar{\chi}_{\nu}\gamma_{\beta}\gamma_{\alpha}\tilde
D_{\rho}\chi_{\sigma}) +e\bar\chi_{\mu} J^{\mu},
\end{eqnarray}
where we added a coupling term to an unspecified supercurrent. At
linear order in the Rarita-Schwinger field $\chi_{\mu}$ the field
equation reads
\begin{eqnarray}
2i\epsilon^{\mu\alpha\nu}\epsilon^{\beta\rho\sigma}\gamma_{\beta}\gamma_{\alpha}
\partial_{\nu}\partial_{\rho}\chi_{\sigma}=-J^{\mu},
\end{eqnarray}
which looks rather non-standard.

To start with we use the $Q$ supersymmetry transformation rule,
$\delta \chi_{\mu}=\partial_{\mu}\epsilon$ to set $\chi_{-}=0$. Then
we would like to go on and use the $S$ superconformal
transformations $\delta \chi_{\mu}=\gamma_{\mu}\eta$ to set one of
the remaining two field components of $\chi_{\mu}$ to zero. However,
this is in direct conflict with $\chi_{-}=0$ since $\chi_{-}$ will
be affected by the superconformal transformation. This means that we
have to design a new superconformal transformation $S'$ that does
not have this problem. We define
\begin{eqnarray}
S'(\eta)=S(\eta)+Q(\epsilon=-\tfrac{\gamma_-}{\partial_-}\eta)=\gamma_{\mu}
\eta-\partial_{\mu}(\tfrac{\gamma_-}{\partial_-}\eta),
\end{eqnarray}
which satisfies $\delta_S' \chi_{-}=\gamma_-
\eta+\partial_-(-\tfrac{\gamma_-}{\partial_-}\eta)=0$. We also find
that
\begin{eqnarray}
\delta_S' \chi_{+}&=&\gamma_+
\eta-(\tfrac{\gamma_-}{\partial_-}\partial_+\eta)\cr \delta_S'
\chi&=&\gamma \eta-(\tfrac{\gamma_-}{\partial_-}\partial\eta).
\end{eqnarray}

Thus we see that the $P^-$ projected parts of $\chi_+$ and $\chi$
can both be set to zero, using the first and second spin component
of $\eta$, respectively\footnote{This is of course restricted to
$p^+\neq 0$, and for $\chi$ in addition to $p^+\neq -\sqrt{2} p$.} .
So the gauge choices we will use are
\begin{eqnarray}
\chi_-=P^-\chi_+=P^-\chi=0.
\end{eqnarray}

Next we would like to implement these conditions in the field
equation and analyze the resulting equations. The covariant
linearized equations are
\begin{eqnarray}
\partial^{\alpha}\partial_{\alpha}\chi_{\mu}-\partial_{\mu}(\partial^{\nu}\chi_{\nu})+\epsilon_{\mu}{}^{\nu\rho}
(\gamma^{\alpha}\partial_{\alpha})\partial_{\nu}\chi_{\rho}=-iJ_{\mu}.
\end{eqnarray}
Using the gauge conditions inside the two expressions in parenthesis
these equations become
\begin{eqnarray}
\partial^{\alpha}\partial_{\alpha}\chi_{\mu}-\partial_{\mu}(-\partial_-\chi_{+}+\partial\chi)+
\epsilon_{\mu}{}^{\nu\rho}
(\gamma^{-}\partial_{-}+\gamma\partial)\partial_{\nu}\chi_{\rho}=-iJ_{\mu}.
\end{eqnarray}

Then the $\mu=-$ component of this equation reads
\begin{eqnarray}
-\partial_{-}(-\partial_-\chi_{+}+\partial\chi) -
(\gamma^{-}\partial_{-}+\gamma\partial)(\partial_{+}\chi-\partial\chi_{+})=-iJ_{-},
\end{eqnarray}
which if further projected with $P^+$ and $P^-$ gives the two
equations
\begin{eqnarray}
\partial_{-}^2\chi_{+}-\partial_-\partial\chi+ -
\gamma\partial\partial_{-}\chi&=&-iP^+J_{-},
\end{eqnarray}
\begin{eqnarray}
\partial_{-}^2\chi&=&\tfrac{i}{2}\gamma^+P^-J_{-}.
\end{eqnarray}
Similarly, the $\mu=+$ component is
\begin{eqnarray}
\partial^{\alpha}\partial_{\alpha}\chi_{+}-\partial_{+}(-\partial_-\chi_{+}+\partial\chi)+
(\gamma^{-}\partial_{-}+\gamma\partial)(\partial_{+}\chi-\partial\chi_{+})=-iJ_{+},
\end{eqnarray}
which splits into the two equations
\begin{eqnarray}
\partial^{\alpha}\partial_{\alpha}\chi_{+}+\partial_{+}\partial_-\chi_{+}-\partial_{+}\partial\chi+
\gamma\partial(\partial_{+}\chi-\partial\chi_{+})=-iP^+J_{+},\cr
\gamma^{-}\partial_{-}(\partial_{+}\chi-\partial\chi_{+})=-iP^-J_{+}.
\end{eqnarray}
Finally, the $\mu=2$ component reads (dropping the index $2$)
\begin{eqnarray}
\partial^{\alpha}\partial_{\alpha}\chi-\partial(-\partial_-\chi_{+}+\partial\chi)+
(\gamma^{-}\partial_{-}+\gamma\partial)\partial_{-}\chi_{+}=-iJ_{\mu},
\end{eqnarray}
and its two component equations are
\begin{eqnarray}
\partial_{-}^2\chi_{+}&=&-\tfrac{i}{2}P^-J,\cr
\partial^{\alpha}\partial_{\alpha}\chi+\partial\partial_-\chi_{+}-\partial^2\chi+
\gamma\partial\partial_{-}\chi_{+}&=&-iP^+J.
\end{eqnarray}

So, we find immediately that the two component fields that remain
after gauge fixing are determined by two of the above equations:
\begin{eqnarray}
\chi_{+}&=&-\tfrac{i}{2\partial_{-}^2}\gamma^+P^-J,\cr
\chi&=&-\tfrac{i}{2\partial_{-}^2}\gamma^+P^-J_-.
\end{eqnarray}
Furthermore, we find from the other component equations conditions
also on $\partial_+\chi$ and $\partial_+\chi_+$, namely
\begin{eqnarray}
\partial_+\chi_{+}&=&-\tfrac{i}{2\partial_{-}}P^+J^+,\cr
\partial_+\chi&=&-\tfrac{i}{2\partial_{-}^2}\gamma^+P^-J_-.
\end{eqnarray}
The last two component equations are just the restrictions on the
supercurrent needed to make it compatible with both local
supersymmetry and local superconformal symmetry. Also the fact that
both field components and their first $\partial_+$ derivative are
determined lead to constraints on the supercurrent. That these
conditions are satisfied by the supercurrents in the topologically
gauged BLG can easily be checked. One explicit example of this will
be presented at the end of the section.

Finally, we turn to the Chern-Simons term for the metric compatible
spin connection. Also here we  have two symmetries to take into
account, the reparameterization invariance and local scale
invariance. As usual, the theory is also locally Lorentz invariant
so we first use this fact to put $\omega_{-\alpha\beta}=0$. This
condition follows also if we use the Lorentz and coordinate
invariances to impose on the dreibein the following constraints at
the linearized level:
\begin{eqnarray}
e_-{}^{-}=1,\,\,e_-{}^{+}=0,\,\,e_-{}^{2}=0,
\end{eqnarray}
and that it is symmetric. Of the remaining dreibein components only
$e_{2}{}^{2}$ is affected by a (linear) local rescaling (after a
light-cone redefinition making it orthogonal to the previous gauge
choices) and thus we can also set
\begin{eqnarray}
e_2{}^{2}=0.
\end{eqnarray}

The spin two Lagrangian is (keeping a spin connection with torsion)
\begin{eqnarray}
L&=&\frac{1}{2}\epsilon^{\mu\nu\rho}
Tr_{\alpha}(\tilde\omega_{\mu}\partial_{\nu}\tilde\omega_{\rho}+
\frac{2}{3}\tilde\omega_{\mu}\tilde\omega_{\nu}\tilde\omega_{\rho})+e
e_{\mu}{}^{\alpha}T^{\mu}{}_{\alpha}.
\end{eqnarray}
To find the field equations we first  vary with respect to the spin
connection which gives
\begin{eqnarray}
\delta
L=-\tfrac{1}{2}\epsilon^{\mu\nu\rho}\delta\tilde\omega_{\mu\alpha\beta}\tilde
R_{\nu\rho}{}^{\alpha\beta} =2\delta\tilde\omega^*_{\mu\alpha}\tilde
R^{**\mu\alpha}.
\end{eqnarray}
Then we need also the dreibein variation of the spin connection
which reads, setting the torsion part to zero,
\begin{eqnarray}
\delta\omega_{\mu\alpha\beta}=e_{[\alpha}{}^{\sigma}
D_{\mid\mu}\delta e_{\sigma\mid\beta]}-e_{[\alpha}{}^{\sigma}
D_{\mid\sigma}\delta
e_{\mu\mid\beta]}+e_{[\alpha}{}^{\sigma}e_{\beta]}{}^{\nu}e_{\mu}{}^{\gamma}
D_{[\nu}\delta e_{\sigma]\gamma}.
\end{eqnarray}
The field equation without torsion is then found to be (see, e.g.
\cite{Horne:1988jf})
\begin{eqnarray}
D_{\rho}W_{\mu\alpha}-D_{\mu}W_{\rho\alpha}=-\tfrac{1}{2}\epsilon_{\rho\mu\sigma}T^{\sigma}{}_{\alpha},
\end{eqnarray}
where the LHS is related to the Cotton tensor
$C_{\mu\nu}={\epsilon_{\mu}}^{\rho\sigma}D_{\rho}W_{\sigma\nu}$ and
\begin{eqnarray}
W_{\mu\alpha}=R_{\mu\alpha}-\tfrac{1}{4}e_{\mu\alpha}R.
\end{eqnarray}
However, we will need only its linearized version:
\begin{eqnarray}
&-&\partial^{\alpha}\partial_{\alpha}\partial_{[\rho}h_{\mu]\nu}+
\partial_{\nu}\partial^{\sigma}\partial_{[\rho}h_{\mu]\sigma}\cr &-&
\tfrac{1}{2}\eta_{\nu[\rho}(\partial^{\alpha}\partial_{\alpha}\partial_{\mu]}h_{\sigma}{}^{\sigma}-
\partial_{\mu]}\partial^{\alpha}\partial^{\beta}h_{\alpha\beta})\cr
&=&-\tfrac{1}{2}\epsilon_{\rho\mu\sigma}T^{\sigma}{}_{\nu}.
\end{eqnarray}

The next step is to implement the above set of gauge conditions on
the linearized dreibein $h_{\mu\nu}$.  From now on we denote the
remaining fields (up to a factor 2) by $h_{++}$ and $h_{+2}$ since
they are really metric components. With these gauge choices we also
find that the trace of $h$ vanishes, and that the field equations
simplify to
\begin{eqnarray}
&-&2\partial^{\alpha}\partial_{\alpha}\partial_{[\rho}h_{\mu]\nu}-2
\partial_{-}\partial^{\nu}\partial_{[\rho}h_{\mu]+}+2\partial_{\nu}\partial^{2}\partial_{[\rho}h_{\mu]2}\cr
&&+ \eta_{\nu [\rho}\partial_{\mu]}(\partial_{-}^2 h_{++}-2
\partial_{-}\partial_{2}h_{+2})\cr &=&-\epsilon_{\rho\mu\sigma}T^{\sigma}{}_{\nu}.
\end{eqnarray}
Reading off the equations for each possible index combination we
find nine equations which imply that the stress tensor is symmetric
and traceless. The remaining five equations can be solved for
$h_{++}$ and $h_{+2}$ and $\partial_+$ derivatives on them in terms
of some expressions involving only the stress tensor, thus giving a
picture similar to the one obtained above for the Rarita-Schwinger
field. Explicitly we find
\begin{eqnarray}
\partial_-^3 h_{++}&=&-2 T_{2-},\cr
\partial_-^3 h_{+2}&=&- T_{--},\cr
\partial_-^2(\partial_+h_{+2})-\tfrac{1}{2}\partial_-^2\partial_2h_{++}&=&\tfrac{1}{2}T_{22},\cr
\partial_-^2(\partial_+h_{++})-2\partial_-\partial_2^2h_{++}+4\partial_-(\partial_+h_{+2})&=&2T_{2+},\cr
\partial_-(\partial_+^2h_{+2})-\partial_2^2(\partial_+h_{+2})+\partial_2^2(\partial_+h_{+2})&=&-T_{++}.
\end{eqnarray}
The solution is
\begin{eqnarray}
h_{++}&=&-2 \partial_-^{-3} T_{2-},\cr
h_{+2}&=&-\partial_-^{-3}T_{--},\cr
\partial_+h_{++}&=&2\partial_-^{-2}T_{2+}-2\partial_-^{-3}(\partial_2T_{22}),\cr
\partial_+h_{+2}&=&\tfrac{1}{2}\partial_-^{-2}T_{22}-\partial_-^{-3}(\partial_2T_{2-}),\cr
\partial_+^2h_{+2}&=&-\partial_-^{-1}T_{++}+\partial_-^{-3}(\partial_2^2T_{22}).
\end{eqnarray}

Finally, as promised, we check that the equations above involving
extra $\partial_+$'s are consistent with expected properties of the
stress tensor. Consider for instance the stress tensor for a free
scalar field. The locally scale invariant action for a scalar field
(in three dimensions) is
\begin{eqnarray}
L=-\tfrac{1}{2}e(g^{\mu\nu}\partial_{\mu}\phi\partial_{\nu}\phi+\tfrac{R}{8}\phi^2).
\end{eqnarray}
This Lagrangian leads to a stress tensor that is traceless and
satisfies the $\partial_+$ conditions mentioned above. The curvature
term turning the free scalar into a conformal theory is present also
in the topologically gauged BLG derived in \cite{Gran:2008qx}.

\section{Conclusions and comments}

Light-cone techniques are often used to keep track of the physical
propagating degrees of freedom while making it possible to solve
for, and thus effectively eliminate, all unphysical local modes. In
this paper this is done for the three-dimensional superconformal BLG
theory describing a pair of M2 branes. The result is an action
expressed completely in terms of the physical light-cone modes
$A,C_{mn},\chi_m$. We also take a first step towards a light-cone
superspace formulation in the spirit of previous work on $\mathcal
N=4$ SYM in four dimensions
\cite{Brink:1982wv,Mandelstam:1982cb,Brink:1982pd} by giving the
kinetic term and a possible answer for the quartic term in
superspace. Some issues related to the sixth order terms are also
discussed.

We then apply these methods to the topologically gauged BLG theory,
discussed recently in \cite{Gran:2008qx}, in order to find out
exactly how the physical modes are embedded into such a topological
higher derivative theory. In spite of the many derivatives it is
possible to solve for all components of the gauge fields (the
dreibein, Rarita-Schwinger and R-symmetry gauge potential) in terms
of BLG matter fields here represented by general currents. However,
one also finds expressions which provide relations between
$\partial_+$ derivatives acting on these field components and the
currents. These latter equations generate constraints on the
currents which are shown  to be satisfied in the simple example of a
conformally coupled scalar field.

We are in this paper dealing with two theories based on Chern-Simons
terms, in the BLG case for ordinary Yang-Mills fields and in the
gauged case also for the gravitational fields in $\mathcal N=8$
superconformal gravity, none of which have any propagating physical
degrees of freedom. Therefore, it would be interesting to  study the
global modes of these topological theories.  Results in this
direction are well-known in the ungauged cases, and in particular
for ABJM theories (see, e.g., \cite{Drukker:2008jm} and references
therein). An analysis of the global modes is easier done using
methods not related to the light-cone gauge, but it may not be
impossible (e.g., a continuation to Euclidean signature can be
performed as explained in \cite{Mandelstam:1982cb}). In fact, it
would be interesting to know to what extent, if at all, global modes
can be studied in the light-cone gauge. Topological aspects of the
gravitational Chern-Simons theory have also been discussed some time
ago in \cite{Horne:1988jf} using a reformulation in terms of an
$SO(3,2)$ ordinary Chern-Simons theory.

The resulting light-cone Lagrangian obtained in this paper rely on a
proper definition of the inverse operator $(\partial_-)^{-1}$ (see,
e.g., \cite{Mandelstam:1982cb}) and avoids the need for higher
powers of it. Such higher powers  do, however, appear in light-cone
discussions of $\mathcal N=8$ four-dimensional gravity
\cite{Brink:1982pd} and may be defined by repeated use of the
definition for a single inverse operator. If the results obtained
here have any bearing on the more complicated gravity theories
remains to be seen.

\acknowledgments

We are grateful to Lars Brink, Martin Cederwall, Gabriele Ferretti,
Ulf Gran and Christoffer Petersson for discussions and especially to
Sung-Soo Kim for a critical reading of the manuscript.

\end{document}